# B

# Big Data and Privacy Issues for Connected Vehicles in Intelligent Transportation Systems

Adnan Mahmood[1,2,3], Hushairi Zen[2], and Shadi M. S. Hilles[3]
[1]Telecommunication Software & Systems Group, Waterford Institute of Technology, Waterford, Ireland
[2]Universiti Malaysia Sarawak, Kota Samarahan, Sarawak, Malaysia
[3]Al-Madinah International University, Shah Alam, Selangor, Malaysia

## Definitions

Intelligent Transportation System is regarded as a smart application encompassing the promising features of sensing, analysis, control, and communication technologies in order to improve the safety, reliability, mobility, and efficiency of ground transportation.

Big Data is an evolving and emerging terminology describing a voluminous amount (*terabytes, petabytes, exabytes, or even zettabyte*) of structured, unstructured, and semi-structured data which could be mined for information.

## Overview

The evolution of Big Data in large-scale Internet-of-Vehicles has brought forward unprecedented opportunities for a unified management of the transportation sector, and for devising smart Intelligent Transportation Systems. Nevertheless, such form of frequent heterogeneous data collection between the vehicles and numerous applications platforms via diverse radio access technologies has led to a number of security and privacy attacks, and accordingly demands for a 'secure data collection' in such architectures. In this respect, this chapter is primarily an effort to highlight the said challenge to the readers, and to subsequently propose some security requirements and a basic system model for secure Big Data collection in Internet-of-Vehicles. Open research challenges and future directions have also been deliberated.

## Introduction

Over the past few decades, a rapid increase in the number of vehicles (both passenger and commercial) across the globe has pushed the present-day transportation sector to its limits. This has thus caused the transportation systems to become highly ineffective and relatively expensive to maintain and upgrade overtime (Contreras et al. 2017). According to one recent estimate (British Petroleum 2017), the number of vehicles on the road has already surpassed 0.9 billion and is likely to be doubled by the end of year 2035. This massive increase in the number of vehicles has not only led to extreme traffic congestion(s) in dense urban environments and hinders economic growth in several ways but also transpires in a number of road fatalities. The World Health Organization (2017) estimates that around 1.25 million people across the world die each year and millions more get injured as a result of road accidents, with nearly half of them being vulnerable road users, i.e., pedestrians, roller skaters, and motorcyclists. Approximately 90% of these road accidents occurs in low- and middle-income economies as a result of their poor infrastructures. From a network management perspective, this results in serious deterioration of quality of service for numerous safety-critical and non-safety (i.e., infotainment) applications and in degradation of quality of experience of vehicular users. Thus, there exists a significant room for improvement in the existing transportation systems in terms of enhancing safety, non-safety (i.e., infotainment), and efficiency aspects; and if properly addressed, this would ultimately set tone for highly efficacious intelligent transportation systems indispensable for realizing the vision of connected vehicles.

The paradigm of vehicular communication has been studied and thoroughly analyzed over the past two decades (Ahmed et al. 2017). It has rather evolved from traditional vehicle-to-infrastructure (V2I) communication to more recent vehicle-to-vehicle (V2V) communication and vehicle-to-pedestrian (V2P) communication, thus laying the foundations for the futuristic notion of vehicle-to-everything (V2X) communication (Seo et al. 2016). While cellular networks have been mostly relied upon for V2V communication due to its extended communication range (i.e., theoretically up to 100 km) and extremely high data rates, its high cost and low reliability in terms of guaranteeing stringent delay requirements do not make it as an ideal mode of communication in highly dynamic networks. On the contrary, the evolution of dedicated short-range communication (DSRC) as a two-way short-to-medium range wireless communication for safety-based V2V applications has been a subject of attention in recent years for engineers and scientists as a result of its efficacious real-time information exchange between the vehicles. While a number of challenges still hinder the effective deployment of DSRC-based vehicular networks, the US Department of Transportation (US DoT) is actively pursuing the same as one of its major research priority for various public safety and traffic management applications including, but not limited to, forward collision warnings, blind intersection collision mitigation, (approaching) emergency vehicle warnings, lane change assistance, and for anticipated traffic and travel conditions (Lu et al. 2014). A topological diagram of next-generation heterogeneous vehicular networks is depicted in Fig. 1.

In addition to both cellular networks and DSRC, numerous other radio access technologies (RATs) are currently being explored for vehicular communication. Since the number of sensors deployed onboard the vehicles (and on roadside infrastructure) is anticipated to increase by manifold primarily due to the introduction of connected and autonomous vehicles, it is necessary to deploy a wireless communication system with the ability to transmit a large amount of data at high data rates. Since DSRC typically allows vehicles to transmit data up to a range of 1 km and with a practical data rate of 2–6 Mbps, and while data rates for cellular networks are also limited to 100 Mbps in high mobility scenarios, the ITS community has also gained momentum for exploring the possibility of millimeter wave communication (mmWave) for vehicular networking. mmWave is expected to offer gigabit-per-second data rate; however, issues such as efficacious channel modeling, security, and beam alignment still pose a considerable challenge to its implementation (Va et al. 2016). Further higher-frequency bands in the frequency spectrum are also gaining attention, and terahertz communication for vehicular networks is also on the cards of researchers in academia and wireless industry, thus paving the way for fifth-generation (5G) and beyond fifth-generation (beyond 5G) wireless networking technologies (Mumtaz et al. 2017).

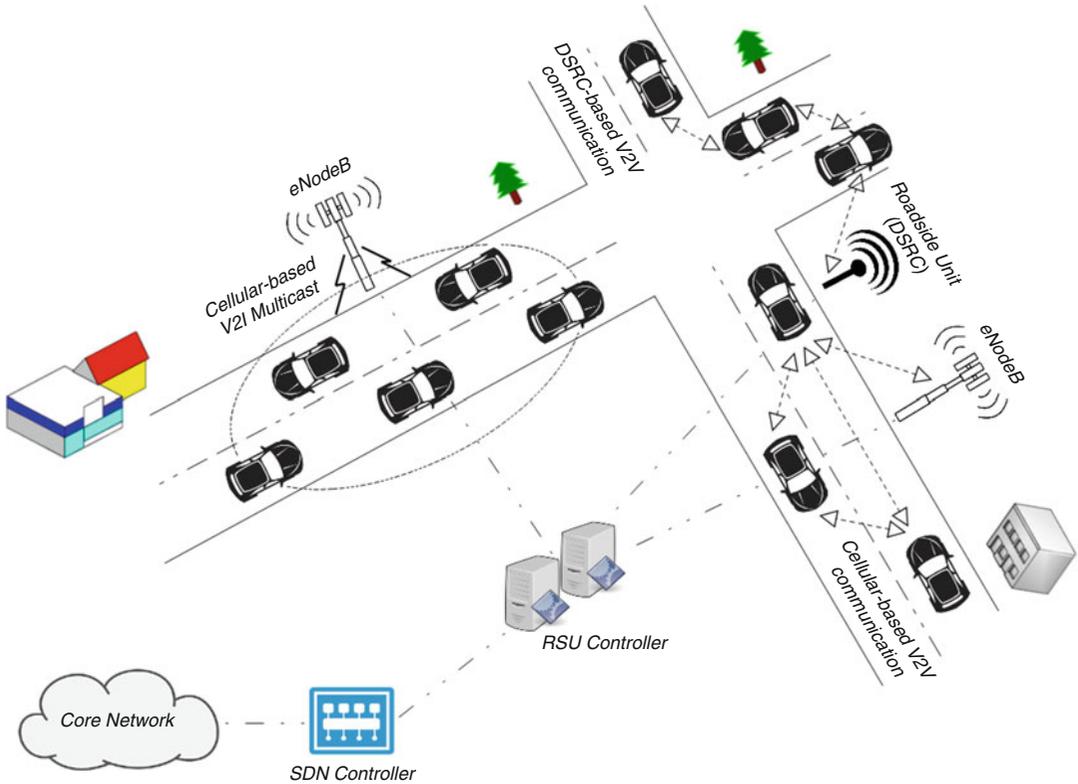

**Big Data and Privacy Issues for Connected Vehicles in Intelligent Transportation Systems, Fig. 1**
Heterogeneous vehicular networking architecture

However, none of these aforementioned technologies possess a potential to fully realize the breadth of vehicular safety and non-safety applications simultaneously and especially when their requirements are in conflict with one another. The ideal approach thus is to ensure a synergy of several RATs so as to provide an appropriate heterogeneous vehicular networking platform in order to meet the stringent communication requirements. Heterogeneity is an important and timely topic but beyond the scope of this work. Since a vehicular network encompasses numerous heterogeneous data sources (i.e., not only from heterogeneous RATs but also from diverse sort of sensing equipment), the amount of data generated is not only huge but also poses a significant challenge to the network's security and stability. Therefore, this chapter, in contrast to other commonly published surveys and chapters, primarily focuses on secure Big Data collection in vehicular networks (interchangeably, referred to as *Internet of Vehicles* in this chapter).

## Toward Big Data Collection in Vehicular Networks

In the near future, several substantial changes are anticipated in the transportation industry especially in light of the emerging paradigms of Internet of Things (IoTs), cloud computing, edge and/or fog computing, software-defined networking, and more recently Named Data Networking, among many others. Such promising notions have undoubtedly strengthened and contributed to the industrial efforts of devising intelligently connected vehicles for developing safe and convenient road environments. Furthermore, with the rapid evolution of high-speed mobile Internet access and the demand for a seamless ubiquitous communication at a much affordable price, new vehicular applications/services are emerging. The European Commission estimates that around 50–100 billion smart devices would be directly connected to the Internet by the end of year 2019, and approximately 507.5 ZB/year of data would be produced due to the same (Sun and Ansari 2016). Moreover, today the smart vehicles have emerged as a multisensor platform, as the usual number of intelligent sensors installed on a vehicle is typically around 100, and this number is anticipated to increase to 200 by year 2020 (Choi et al. 2016). The data streams generated via these sensors are not only high in volume but also possess a fast velocity due to highly dynamic nature of vehicular networks. Moreover, information generated from numerous social networking platforms (i.e., Weibo, Facebook, Twitter, WhatsApp, WeChat) are regularly accessed by the vehicular users and generate a lot of traffic data in real time. This data is spatiotemporal in nature due to its dependence on time and geographical location of vehicles. Also, since the trajectory of vehicles is usually reliant on the road distribution across a huge geographical region, the collected information is also heterogeneous, multimodal, and multidimensional and varies in its quality. This context is pushing the concept of traditional vehicular ad hoc networks to large-scale Internet of Vehicles (IoVs), and all of this collected data converges as Big Data in vehicular networks which ultimately is passed via the core networks to both regional and centralized clouds.

Nevertheless, today's Internet architecture is not yet scalable and is quite inefficient to handle such a massive amount of IoV Big Data. Also, transferring of such data is relatively time-consuming and expensive and requires a massive amount of bandwidth and energy. Moreover, real-time processing of this collected information is indispensable for a number of safety-critical applications and hence demands for a highly efficient data processing architecture with a lot of computational power. As vehicular networks are highly distributive in nature, a distributed edge-based processing is recommended over traditional centralized architectures. However, such distributed architectures mostly have limitations of compute and storage and therefore are unable to cache and process a huge amount of information. Last but not the least, it is quite significant to design a secure mechanism which ensures that collection of IoV Big Data is trusted and not tampered with. There is a huge risk of fraudulent messages injected by a malicious vehicle that could easily endanger the whole traffic system(s) or could potentially employ the entire network to pursue any dangerous activity for its own wicked benefits. Thus, *how to effectively secure the Big Data collection in IoVs deserves researching*. In this context, this chapter is of high importance so as to bring to forth the significance of such challenges for the attention of academic and industrial research community.

## Security Requirements and System Model for Secure Big Data Collection in IoVs

Big Data primarily has three main characteristics, (a) volume, (b) variety, and (c) velocity, and this is also referred to as 3Vs (Guo et al. 2017). Vehicles usually collect a massive amount of data from diverse geographical areas and with various heterogeneous attributes, thus ultimately converging to IoV Big Data with variation in its size, volume, and dimensionality. The analytics of such Big Data can help network operators to optimize the overall resource(s) planning of next-generation vehicular networks. It would also facilitate the national transportation agencies to critically analyze and subsequently mitigate the dense traffic problems in an efficient manner, in turn making the lives of millions of people comfortable and convenient. These are some of the potential advantages that have lately encouraged the vehicles' manufacturers to build large-scale Big Data platforms for the intelligent transportation systems. However, all of this data analytics and subsequent decision making becomes impractical if any malicious vehicle is able to inject data in the IoV Big Data stream which would have severe implications for both safety and non-safety vehicular applications. In terms of safety applications, any maliciously fed infor-mation may lead to false trajectory predictions and wrong estimation of a vehicle's neighboring information in the near future which could prove quite fatal for passengers traveling in both semi-autonomous and fully autonomous vehicles. In terms of non-safety applications, this could not only lead to a considerable delay in the requested services but also exposes a user's privacy data to extreme vulnerabilities (Singh et al. 2016; Zheng et al. 2015). Thus, any secure IoV information collection scheme should adhere to the following requirements so as to guarantee the secure Big Data collection:

*Data Authentication* – to verify the identities of the vehicles or vehicular clouds (i.e., vehicles with similar objectives and in vicinity of one another form vehicular clouds for resource-sharing purposes)

*Data Integrity* – to ensure that the transmitted data has been delivered correctly without any modification or destruction, as this would ultimately become part of the Big Data stream

*Applications/Services Access Control* – to warrant that each respective vehicle has access to the applications/services it is only entitled for

*Data Confidentiality* – to guarantee a secure communication among vehicles participating in a networking environment

*Data Non-repudiation* – to ensure that any particular vehicle could not deny the authenticity of another vehicle

*Anti-jamming* – to prevent intrusion attempts from malicious vehicles which may partially or completely choke the entire network

An ideal approach to tackle these requirements would be via a hierarchical edge-computing architecture as depicted in Fig. 2. Each vehicle (or a vehicular cluster), along with their respective vehicular users, is associated with their private proxy VMs situated in the edge node which not only collects the (raw) data streams from their registered network entities via their respective base stations or access points but also classifies them into various groups depending on the type of data and subsequently generates metadata which is passed to regional/centralized IoV Big Data centers. The metadata encompasses certain valuable information, i.e., geographical locations of vehicles and corresponding timestamps, type of vehicular applications accessed or infotainment services and contents requested, and a pattern recognition identifier that checks for any irregularities in data stream against various prescribed network operator policies and also by comparing it with previously known patterns. In this way, a suspicious content could be filtered out at the edge without passing it to the regional/centralized storage. The metadata only contains valuable information without violating a user's privacy. Also, as several vehicles and vehicular users requests similar type of applications/services and pass identical information about their surroundings (i.e., provided if they are in vicinity of one another), intelligent content similarity schemes should be employed in order to avoid any duplication that may lead to excessive computational overheads both at the edge and at the regional/centralized cloud level.

Thus, in this way, not only the computational overheads can be considerably reduced but the amount of IoV Big Data that needs to be cached can also be significantly minimized. It is therefore pertinent to mention that (both) instantaneous and historical data is of critical importance in making intelligent decision making in IoV architectures. Nevertheless, it is also not practically possible to store all of the historical IoV Big Data for an infinite time due to storage issues globally. Thus, it is of extreme importance to not only secure the collection of IoV Big Data but to also ensure that the said data streams should not be excessively cached but disposed after a certain time. Our envisaged architecture hence proposes to structure the raw data streams into valuable metadata, which not only consumes less resources in making a requisite decision but also takes less storage space in contrast to raw data streams.

## Open Research Challenges and Future Directions

Unlike traditional IoT entities which are generally static in nature, IoV is a highly dynamic or-

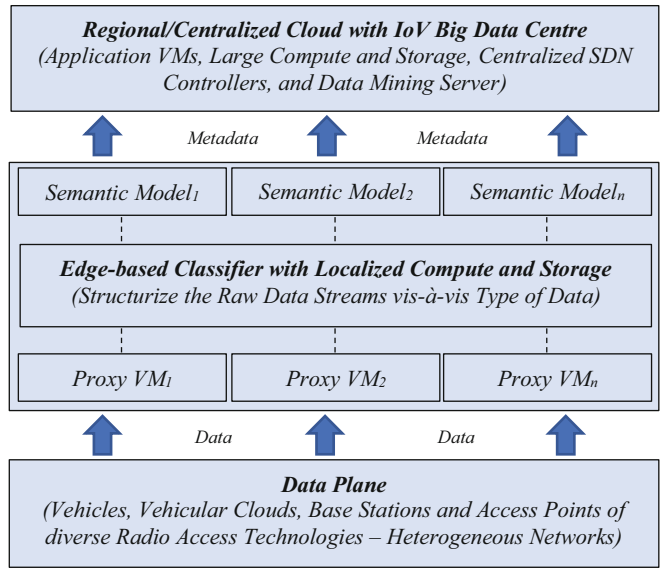

**Big Data and Privacy Issues for Connected Vehicles in Intelligent Transportation Systems, Fig. 2** Edge computing architecture for secure IoV Big Data collection ganism, and this feature makes it quite difficult to effectively deploy the next-generation intelligent transportation system platforms, at least in their current form. Since vehicles usually traverses at quite high geographical speeds, their neighborhood also changes dynamically. Thus, it is challenging to accurately apprehend a vehicle's current and anticipated surrounding neighborhood on run-time basis. A major problem is that by the time the IoV information has been captured, analyzed, and presented to network operators or interested vehicles/vehicular users, the real-time situation has already changed, and the IoV data recently presented has become obsolete. Thus, IoV standalone might not make any essence in the long run; rather, it has to be further optimized with an emerging yet promising paradigm of software-defined networking which through its centralized controllers (possessing a globalized topological view) addresses the trajectory and neighborhood estimation problems to a much greater extent.

The other issue pertains to density of the vehicular networks. If each individual vehicle starts communicating with its edge (i.e., for reporting status information or requesting vehicular applications or infotainment services), this could transpire in a network broadcast storm that may severely deteriorate the overall network performance, and consequently the service-level objectives may not be fulfilled. In this respect, a notion of vehicular cloud has been proposed in order to mitigate the said threat. However, for this purpose, efficacious *vehicular cloud formation schemes* have to be envisaged. Although the concept of vehicular cloud is much similar to that of a cluster formation in sensor networks or more precisely to vehicular platoons in vehicular ad hoc networks, the actual challenge here is again the dynamic nature of vehicular networks due to which vehicular clouds could make and break at a rapid pace resulting in overall wastage of network resources. Also, it is important that no malicious vehicle becomes part of the vehicular cloud, as this could easily become fatal for the entire IoV environment.

Finally, with a recent push for openness of IoV data sharing among diverse cloud service providers, it has become quite imminent that each cloud service provider should protect its infrastructures from a diverse range of malicious attacks by employing specialized access controls, isolation, encryption, and sanitization techniques. Hence, the input data from other cloud service providers need to pass via a validation process so that any malicious, faulty, or compromised data may be timely detected and subsequently blocked in order to protect the data integrity of the whole IoV ecosystem.

## Cross-References

▶ Big Data Analysis for Smart City Applications
▶ Big Data in Smart Cities

## Acknowledgments

The corresponding author would like to acknowledge the generous support of the Ministry of Higher Education, Government of Malaysia, for supporting the said research work through its Malaysian International Scholarship Grant, KPT. 600–4/1/12 JLID 2(8).